# Identification of Biomarkers Controlling Cell Fate In Blood Cell Development

Maryam Nazarieh[1,2], Volkhard Helms[2], Marc P. Hoeppner [1], Andre Franke [1]

1. Institute of Clinical Molecular Biology, Christian-Albrecht-University of Kiel, 24105, Kiel, Germany

2. Center for Bioinformatics, Saarland University, Saarbruecken, Germany

## Abstract
A blood cell lineage consists of several consecutive developmental stages from the pluri- or multipotent stem cell to a state of terminal differentiation. Despite their importance for human biology, the regulatory pathways and gene networks that govern these differentiation processes are not yet fully understood. This is in part due to challenges associated with delineating the interactions between transcription factors (TFs) and their target genes. A possible path forward in this issue is provided by increasingly available expression data as a basis for linking differentiation stages and gene activities. Here, we present a novel hierarchical approach to identify characteristic expression peak patterns that global regulators expose along the differentiation path of cell lineages. Based on such simple patterns, we identify cell state-specific marker genes and extract TFs that likely drive their differentiation. Integration of the mean expression values of stage-specific "key player" genes yields a distinct peaking pattern for each lineage that is used to identify further genes in the dataset behaving similarly. Incorporating the set of TFs which regulate these genes incurred at a set of stage-specific regulators controlling the biological process of cell fate. As proof of concept, we consider two expression datasets covering key differentiation events in blood cell formation of mice.

## Introduction

Cell fate describes a biological program which determines how a less specialized cell type develops into a more specialized one. For each transition out of a particular state, this involves a decision to either self-renewal or differentiate into daughter cells (Garcia-Ojalvo et al., 2012). It is well-accepted that such processes are tightly regulated by transcriptional networks, typically centered around a discrete number of transcription factors (Moignard & Göttgens, 2014). Knowing the "key players" involved in these events may thus not only serve as a predictive marker to help in determining the differentiation stage of cells, but in extension could potentially also serve a clinical purpose, for example by aiding in the search for therapeutic targets across different diseases involving aberrations in the composition of cell types/stages (An et al., 2014).

One of the best-studied tissues is blood, which is already widely used in diagnostics. Especially in complex blood-related diseases such as leukemia, understanding the manifestation of the disease and monitoring its progression and response to treatment could greatly benefit from a deeper understanding of the underlying regulatory processes and key "actors" that govern blood cell differentiation. However, delineating lineage-specific regulatory networks is a challenging task, typically requiring the costly integration of multiple data types – particularly from various "omics" technologies. Previous work using a complex multi-omics approach identified a set of 16 "global regulators" driving the differentiation of blood cells across 6 discrete stages - Embryonic stem cells (ESCs), Mesoderm (MES), Hemangioblast (HB), Hemogenic endothelium (HE), Hematopoietic progenitor (HP) and Macrophages (MAC) (Goode et al., 2016). It is very plausible to assume that these global regulators stand at the top of the regulatory hierarchy and indirectly govern particular cellular identity. Interestingly, although the overall network is preserved across developmental stages, analysis of the characteristic changes in expression (Figure 1) suggests that these "global regulators" contribute differently at various stages.

We have previously developed a method that reconstructs the core components of a regulatory network from gene expression data and defines a so-called "minimum dominating set" (MDS), i.e. the minimum set of TFs that dominate the whole network through their interactions. A modification of this concept is the "minimum connected dominating set" (MCDS) which searches for a minimum number of genes that are connected and control the underlying co-network (Nazarieh 2018; Nazarieh et al. 2016; Nazarieh and Helms 2019). When applied to expression data, one should expect that a key transcription factor being most strongly associated with a certain differentiation stage has a peak expression at that stage relative to the other stages of that lineage. Genes directly regulated by such a key player can be expected to mimic its expression profile, allowing their assignment to a given regulator and cellular stage.

In the present work, we introduce an approach to identify stage-specific key regulators that are likely to control cell fate in a differentiation/developmental or resistance pathway. We demonstrate the usefulness of the approach on the example of two expression data sets that investigated blood cell differentiation in mice (Goode et al, 2016; Bock et al., 2012).

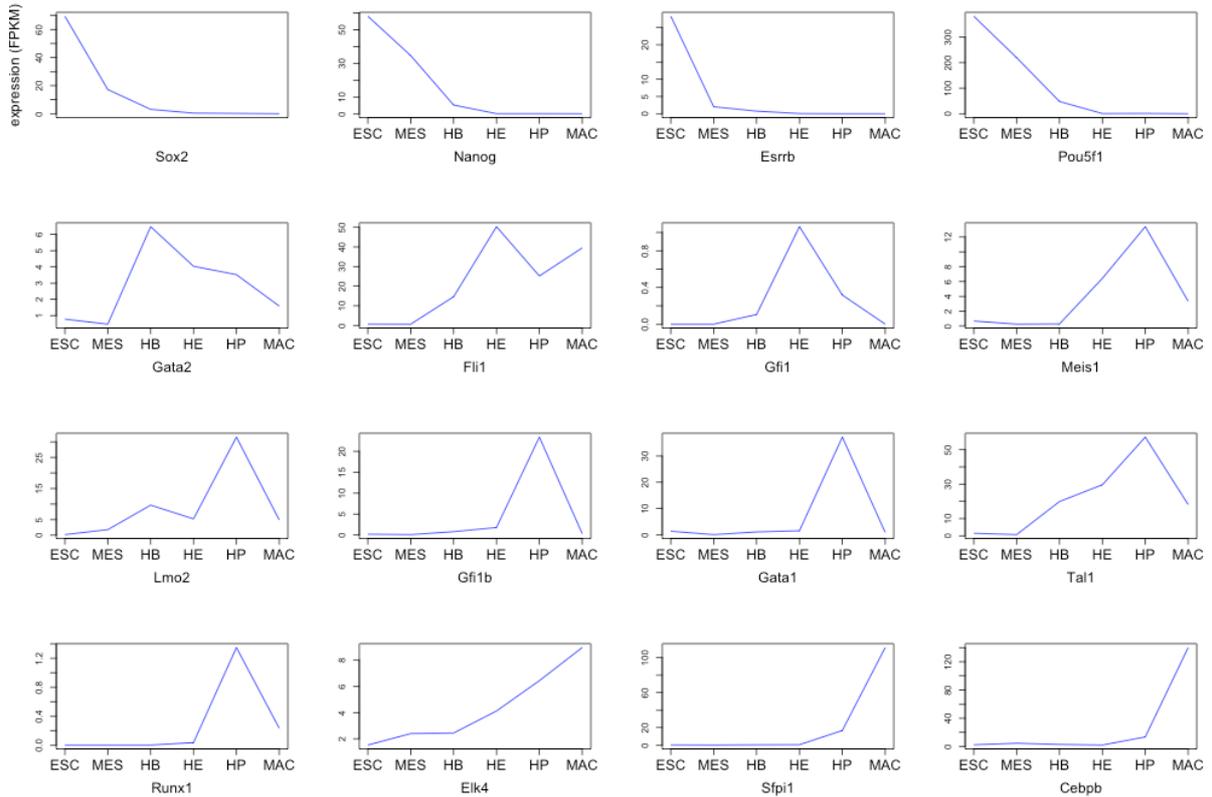

*Figure 1: Expression of the global regulators driving hematopoietic specification for all six stages of blood development starting from ESCs (stage 1) to terminally differentiated macrophages (stage 6) (Goode et al. 2016).*

# Methods

## Overview

Fig. 2 illustrates the workflow of the entire approach. First, we derive diagnostic expression profiles to identify genes that are centrally involved in the cellular differentiation path (Figure 2A). Next, we integrate the expression pattern of cell-specific developmental genes across full individual lineages (Figure 2B). From this, a set of correlated genes and associated TFs is identified (Figure 2C). This preliminary network is further refined by incorporating experimentally validated data e.g. from a TF-gene interaction database such as TRRUST to define a sub regulatory network whose target genes follow the aforementioned expression pattern and have a well-defined TF regulator (Figure 2D). Finally, we present an algorithm that finds the regulatory path that connects the target genes that are tightly regulated by multiple TFs (Figure 2 E). We suggest that the set of target genes and TFs that connect them as most suggested candidates for the cell fate process. A functional enrichment analysis is then used to investigate the biological processes these identified TFs have previously shown to be involved in.

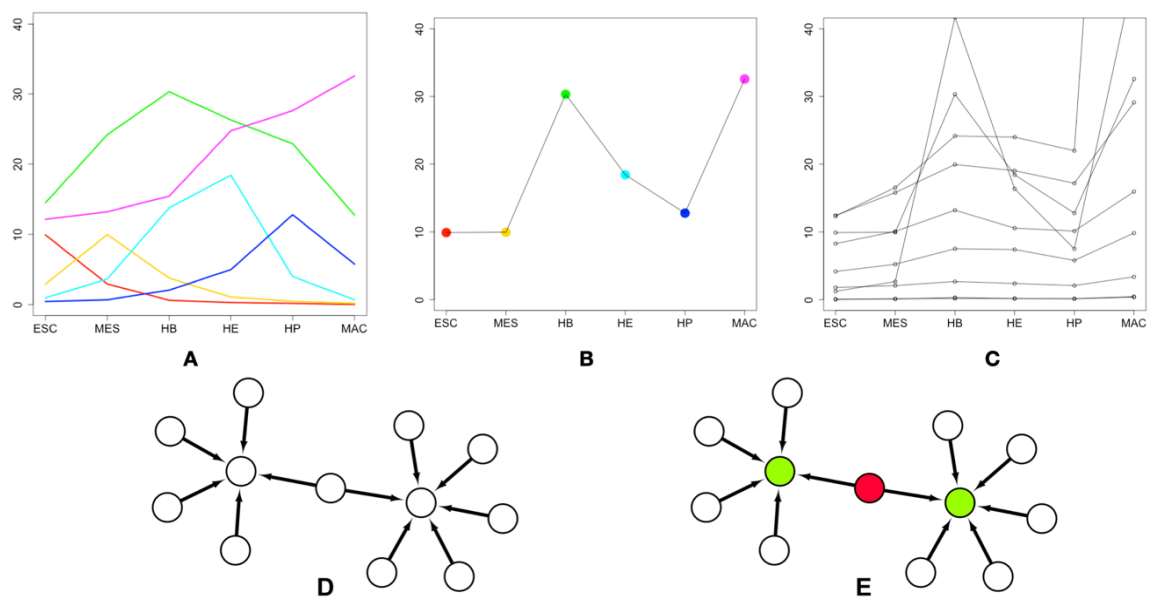

*Figure 2: Overview of how biomarkers are identified that control or drive a development cell fate process (panel A). Fictitious expression profiles (y-axis) of six selected transcription factors (TFs) across six developmental stages (x-axis). (B) TFs are identified having peak expression in the respective stage. This step yields the stage-specific key regulators (TFs). (C) Further genes are identified having highly correlated expression profiles to one of the stage-specific key regulators of panel B (here, one of the TFs peaking in the terminal stage MAC). (D) A gene-regulatory (GRN) network is constructed including all stage-specific key regulators and their correlated target genes. This GRN includes TFs and target genes from all stages. (E) A regulatory pathway is identified (see methods) that connects all "influencer" nodes.*

## Datasets

The first case study was based on genome-wide RNA-seq expression profiles (Goode et al., 2016) in form of FPKM values across six consecutive differentiation stages, namely ESC, MES, HB, EH, HP and MAC (GEO accession GSE69080). The microarray data for the second case study was published by (Bock et al., 2012). As mentioned in that paper, the data were obtained as CEL files and normalized in the same order to reduce batch effects. The data includes 13 cell populations sorted by FACS analysis across 6 lineages.

## Regulatory relationships

Data on the relationship between TFs and their target gene(s) were taken from the TRRUST database v2 (Han et al., 2018) that was compiled based on literature curation. This release of the database includes 6552 TF-target interactions for 828 mouse TFs.

## Workflow: Prioritization of the candidates of the cell fate process

1. Acquisition of stage-specific expression pattern from the tissue-specific global regulators.
2. Identification of genes and TFs that follow a stage-specific expression pattern (First layer candidates) (see panel A of Fig. 2).
3. Integration of stage-specific expression pattern for the cell lines across the lineage (see panel B of Fig. 2).
4. Identification of further genes and TFs that follow the same integrated expression pattern (Second layer candidates) (see panel C of Fig. 2).

5. Identification of the TFs that regulate the candidates in the second layer (Third layer candidates).
6. Reconstruct a network whose regulators are from the third layer and targets are from the second layer (see panel D of Fig. 2).
7. Identification of high-indegree nodes in the network.
8. Find the regulatory pathway that connects the nodes in step 7 (Fourth layer candidates) (see panel E of Fig. 2).

**Randomization Algorithm:**

Input: A set of correlated genes following the integrated pattern of gene expression across the stages in one lineage.
Output: Overlap significance of the correlated genes in the original data and the shuffled ones.

1. Compute the set of correlated genes following the integrated expression pattern based on shuffled data.
2. Compute the overlap between the correlated genes and the correlated genes taken from shuffled data.
3. Report the number of times that the overlap between the correlated genes in the original data and the correlated genes from 1000 shuffled data is greater than 0.05.

**PathDevFate Algorithm: Find the regulatory path that involves a certain set of nodes**

Input: A network that is obtained from step 6 of the above-mentioned pipeline.
Output: A set of genes and TFs with assigned roles of influencers and connectors.

1. Identify the set of nodes that are regulated by at least one TF.
2. Specify a threshold (here denoted by "l") as a measure of in-degree threshold.
3. Select the nodes whose number of incoming edges exceeds "l". These are termed "influencers".
4. Find a path that connects the influencer nodes by adding a minimum number of further (Steiner) nodes ("connectors").

In step 2, a threshold is introduced that provides a balance between the number of influencers with respect to the number of incoming edges and the number of TFs which are supposed to connect them (which depends on the distance these influencers have from each other). This suggestion serves to capture the high-indegree nodes and imposes a minimum number of TFs to the regulatory pathway.

**Functional annotation**

The biological function of the genes in each stage was evaluated using the enrichment analysis tool provided at the DAVID portal of NIH (version 6.8) based on the functional categories in GO Direct (Huang et al., 2009). *p*-values below the threshold of 0.05 obtained by the hypergeometric test were adjusted for multiple testing using the Benjamini & Hochberg (BH) method (Benjamini, 2016).

# Results

The main goal of this study was to derive an approach that identifies a connected set of cell-fate regulating genes. For this, we implemented the hierarchical strategy illustrated in Fig. 2. The first layer includes the stage-specific TFs and genes that are involved in cellular differentiation. The second layer consists of further genes and TFs following the same integrated stage-specific expression pattern. The third layer is formed by those TFs that regulate the candidates in the second layer. This helps to reconstruct a regulatory network from the correlated genes following the integrated expression pattern with a set of TFs that regulate them (fourth layer). Finally, we derive the regulatory path that connects the set of correlated genes that are regulated by multiple TFs (PathDevFate, see Methods). Basically, the target genes that are tightly regulated by multiple transcription factors are flagged as "influencers" and the nodes that connect them as "connectors". As proof of concepts, we applied the method to two datasets of murine blood differentiation. The first case study was a lineage of six stages starting at ESC and leading to MAC (Goode et al., 2016). We then extended the concept by setting rules defined for cellular differentiation in (Artyomov et al., 2010) and applied it to expression data from across 6 murine cell lineages in blood formation (Bock et al., 2012) starting at HSC and leading to either CD4, CD8, B-cells, erythrocytes, granulocytes or monocytes.

## Dataset 1: Differentiation of murine blood stem cells

In published multi-omics data on murine blood stem cells (Goode et al., 2016), we analyzed the expression profiles of a set of "key" transcription factors across the differentiation stages. We then applied these profiles to the full set of expression data to identify genes (and further TFs) having strongly correlated expression patterns. Figure 3 shows the expression pattern of the TFs that were included in the set of "key" TFs. Obviously, multiple TFs have peaks in each of the individual differentiation stages.

This analysis, yielding our "first" gene layer, identified between 197 (HP) and 692 (HB) correlated gene expression profiles. Included in this are between 10 (MAC) and 54 (HB) TFs, such as SOX2 and ESRRB (Table S1). For each stage, we considered the identified genes to reconstruct functional profiles of the correlated genes based on gene ontology terms (GO) (Supplementary tables S2-S5).

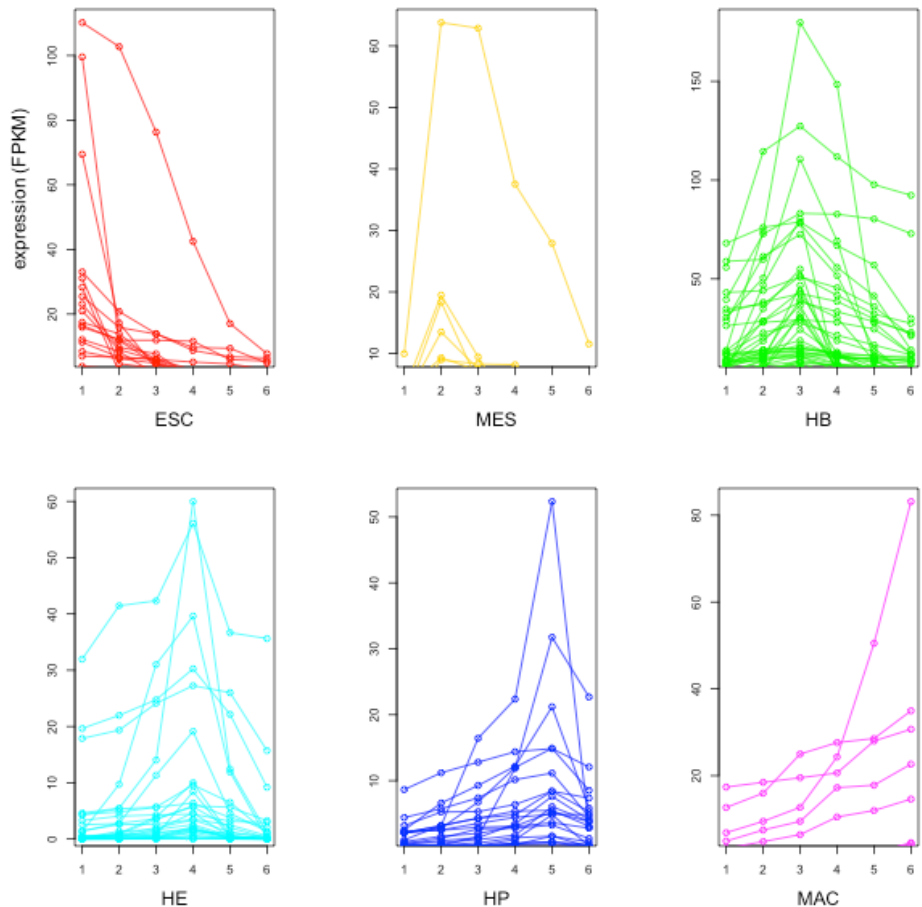

*Figure 3:Expression pattern of identified TFs in six stages of ESC, MES, HB, HE, HP and MAC that follow the global expression pattern.*

In order to understand the molecular mechanisms governing each differentiation stage, we next performed a functional enrichment analysis using both gene ontology (GO) terms and KEGG pathways for the key transcription factors (Supplementary Table S6-S11) found in each differentiation stage as well as for their (known) target genes (Supplementary Table S12-S17), respectively.

Table S6 for ESC includes GO terms such as stem cell differentiation (GO:0048863), multicellular organism development (GO:0007275), endoderm development (GO:0007492) and cell differentiation (GO:0030154), respectively. Moreover, the set of genes (Onecut1, Esrrb, Id1, Sox2, Zic3) belonging to the following KEGG pathway: signaling pathways regulating pluripotency of stem cells (mmu04550) came up. Table S7-S11 list the enriched GO terms and KEGG pathways for the identified TFs in MES, HB, HE, HP and MAC. The lists include further specialized GO terms in addition to some of the aforementioned terms such as patterning of blood vessels (GO:0001569), cell fate commitment (GO:0045165), heart development (GO:0007507) and hemopoiesis (GO:0030097), respectively and also the KEGG pathway: acute myeloid leukemia (mmu05221).

We inferred the set of target genes for the set of "key player" TFs at each developmental stage from the TF-gene interaction database TRRUST (Han et al., 2018). Enrichment analysis for the set of identified target genes in the ESC stage yielded the enriched biological process GO terms listed in Table S12. The list includes GO terms such as proliferation (GO:0042127), multicellular organism development (GO:0007275), stem cell differentiation (GO:0048863), cell differentiation (GO:0030154), cell fate commitment (GO:0045165), cell development (GO:0048468) and cell proliferation (GO:0008283), respectively. Tables S13-S17 list the enriched GO terms for the target genes in other developmental stages. In addition to common GO terms, distinct GO terms such as BMP signaling pathway involved in heart development (GO:0061312) and Wnt signalling pathway (GO:0016055) are added in MES stage. More specialized GO terms appear in later stages HB and HE and HP such as GO:0001889, GO:0002326, GO:0043583 and GO:0001654 with annotations liver development, B cell lineage commitment, ear development and eye development, respectively. Although the TFs identified in each particular developmental stage also follow the aforementioned expression pattern, they expose different expression levels. The histograms in Figure 4 show the frequency of TFs based on their expression level. In general, there are many more TFs with low expression (e.g. 0-10, 0-20 etc.) than with high expression (above 50). There is an initial increase in the absolute number of patterned TF from 13 (ESC), 14 (MES) to 34 (HB), followed by a corresponding decline over 22 (HE), 15 (HP) to 6 (MAC).

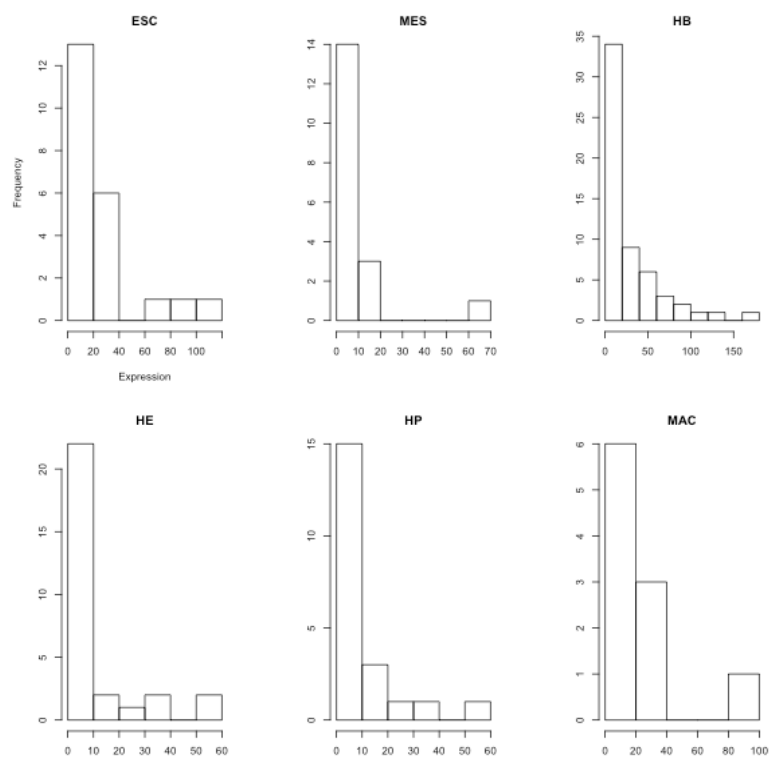

*Figure 4: Histograms of stage-specific TF expression levels in the blood cell lineage show a quasi-exponential decay. E.g. for ESC, 13 TFs have expression levels between 0 and 20, 6 TFs have expression levels between 20 and 40 etc. Only 3 TFs have expression levels above 60. A similar behavior is found for all other developmental stages.*

Genes that act in the same biological processes are expected to (partially) share activity profiles (Huttenhower and Troyanskaya 2008). Thus, in order to identify additional members of a given candidate network, we extracted genes that mimic the same expression patterns exhibited by the previously identified stage-specific genes. To qualify, the expression of these genes had to describe a monotonic relationship (based on the Spearman's rank correlation) with the mean value of stage-specific genes (Figure 5), applying a threshold of greater than 90 percent.

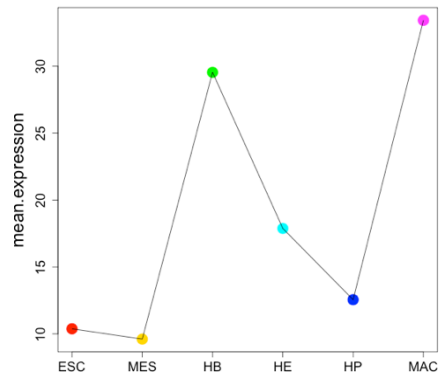

*Figure 5: Depiction of the mean expression of key TFs across six stages of blood cell differentiation (from ESC to MAC).*

Computation of the Spearman's rank correlation (Figure 6) led to 243 genes (Table S18) including 13 TFs with more than 0.9 correlation. These TFs are considered as the candidates in the second layer.

To check the statistical significance of the correlated genes, we resampled the data 1000 times, identified patterned genes in each case, and measured the overlap between the correlated genes in the original data set and those determined from the resampled data, see Figure 7. The overlap was measured based on the Jaccard index as the ratio of intersection between the set of correlated genes and the resampled data over the union of the two sets. Only 3 out of 1000 cases had a similarity higher than 0.05 between the correlated genes in the original data and the correlated genes in the shuffled data (*p*-value of 0.003). Thus, the stage-specific genes identified in the real data are rarely identified based on randomly shuffled data, which strengthens the biological meaningfulness of this analysis.

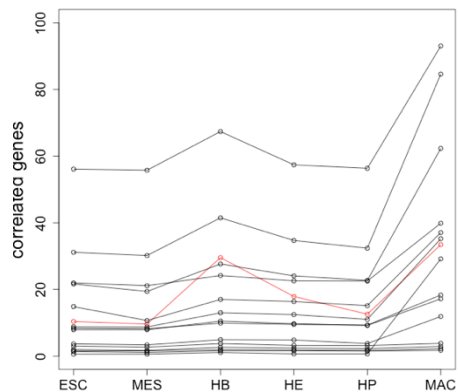

*Figure 6: Depiction of the correlated genes. The red curve shows the pattern of integrated mean expression of all the cells in the lineage. The black curves represent correlated genes that have perfectly positive correlation based on the Spearman method (threshold > 0.9)*.

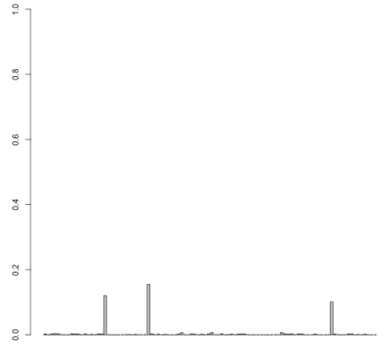

*Figure 7: The number of overlapped genes between the above-mentioned correlated genes and resampled data.*

Next, we sought to identify known regulators of this initial set of co-expressed genes using data from the TRRUST database. This analysis found 83 TFs interactions, which then form the third layer of our analysis (Table S19). The intersection with cell-specific TFs of ESC, MES, HB, HE, HP and MAC identified in the first layer includes (Etv4, Hdac1, Prdm16, Sox2), (Foxo4), (Atf2, Etv2, Gata4, Msx2, Snail1), (Ebf1, Smad3), (Stat5a, Stat5b, Thra), (Arid3a, Stat5b), respectively. All these genes were previously reported to have specific roles in cell fate commitment (Ackermann et al., 2011; Avilion et al., 2003; Babajko et al., 2015; Beuling et al., 2008; Bourgeois & Madl, 2018; Dunn et al., 2004; Garg et al., 2018; Horvay et al., 2015; P. Liu et al., 2015; X. Liu et al., 1996; Rhee et al., 2014; Zandi et al., 2008).

Table S20 shows the functional enrichment analysis (biological process) and KEGG pathways for the 83 TFs with p-values, using a hypergeometric test and adjusted for multiple testing using the Benjamini & Hochberg (BH) method (Benjamini, 2016) below a threshold of $p \leq 0.05$. Notable GO terms on this list include: GO:0008285 negative regulation of cell proliferation, GO:0008284 positive regulation of cell proliferation GO:0043066 negative regulation of apoptotic process, GO:0043065 positive regulation of apoptotic process, GO:0002360 T cell lineage commitment, GO:1902262 apoptotic process involved in patterning of blood vessels, GO:0048863 stem cell differentiation, GO:0030154 cell differentiation, GO:0007507 heart development, GO:0007275 multicellular organism development, GO:0033077 T cell differentiation in thymus, GO:0030217 T cell differentiation

Finally, using information from the TRRUST database, a regulatory network was reconstructed whose nodes are confined to the candidates of the second and third layer. The networks demonstrate the connectivity between the candidates in the second and third layer. The number of TFs in the network exceeds the number of target genes (Table S20) so that the network contains few genes with a high number of incoming edges. In the network having 90 interactions, the 83 regulators were taken from the third layer and 21 target genes from the second layer. This network contains three high-indegree nodes such as (Ccnd2, Pparg and Ihh) in the largest connected component that are connected through Masx2 and Foxo1, see Figure 8.

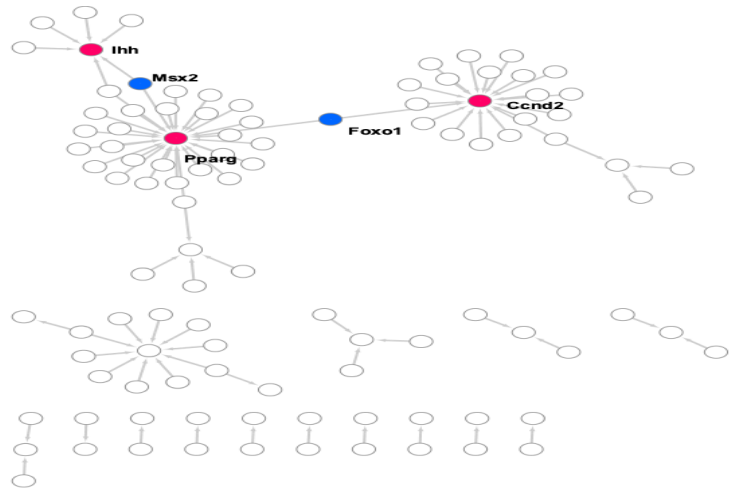

*Figure 8: TF-target network for the set of correlated genes derived from TRRUST database. Influencers (red nodes) which are the stage-specific target genes that are regulated by more than 5 TFs and connectors are TFs (blue nodes).*

## Dataset 2: Blood Stem Cell Differentiation along multiple lineages (Bock et al. 2012)

The previous section focused on a single example of cellular differentiation in blood formation, starting from previously characterized "key" transcription factors. Therefore, we next expanded our initial approach and applied our concept of "key" expression profiles to a more complex dataset, consisting of 6 differentiation lineages starting at mouse blood stem cells (Bock et al, 2012). Differentiation of these lineages was shown by the authors to follow a gradual path of changing expression profiles through up to 6 steps into a fully differentiated cell (Figure 9). To derive the developmental genes and TFs we not only relied on the cell-specific expression pattern as outlined above, but also exploited the computational model and the rules suggested by (Artyomov et al., 2010). Under this model, each cell is defined by two expression and epigenetic states network layers. There are a set of master regulators that define the cellular identity. On the event of cellular differentiation, the activated gene module suppresses the activity of the competitor cells either in relationship of parent cell or daughter branch cells. Here, we modified the rules to the extent that developmental regulators specific to each cell have superiority in terms of gene expression level to the competitor neighboring cells while following the cell-specific expression pattern from the top of the hierarchy until terminally differentiated cells.

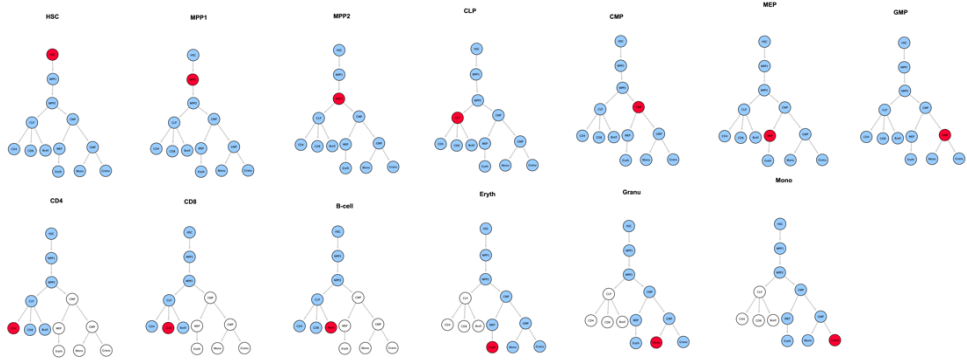

Figure 9: *Red colored nodes denote the gene modules whose expression pattern are the highest among all the cells in the blood differentiation. Blue color nodes stand for the genes whose expression pattern are lower than the red color nodes. The parent nodes before the red color node have gradual increase in the expression pattern and the daughter blue nodes have the gradual decrease which reaches to the minimum possible at the terminally differentiated cells. White color nodes are the cells whose expression level are not considered.*

The aforementioned patterns led to the identification of between 4 and 128 cell-stage specific genes for the different cell types under consideration (Supplementary Table S21), including several well-known TFs.

Figure 10 represents the changes of mean expression value of constituent cells along the cell lineages starting from HSC until a terminally differentiated cell type (e.g. CD4, CD8, B cell, Erythrocyte (Eryth), Granulocyte (Granu) or Monocyte (Mono)). The stage-specific genes of erythrocytes and granulocytes have particularly high expression levels in the terminally differentiated stage. For CD4, CD8, B-cells, and monocytes, an inverse trend is observed. The biological relevance of this is presently unclear.

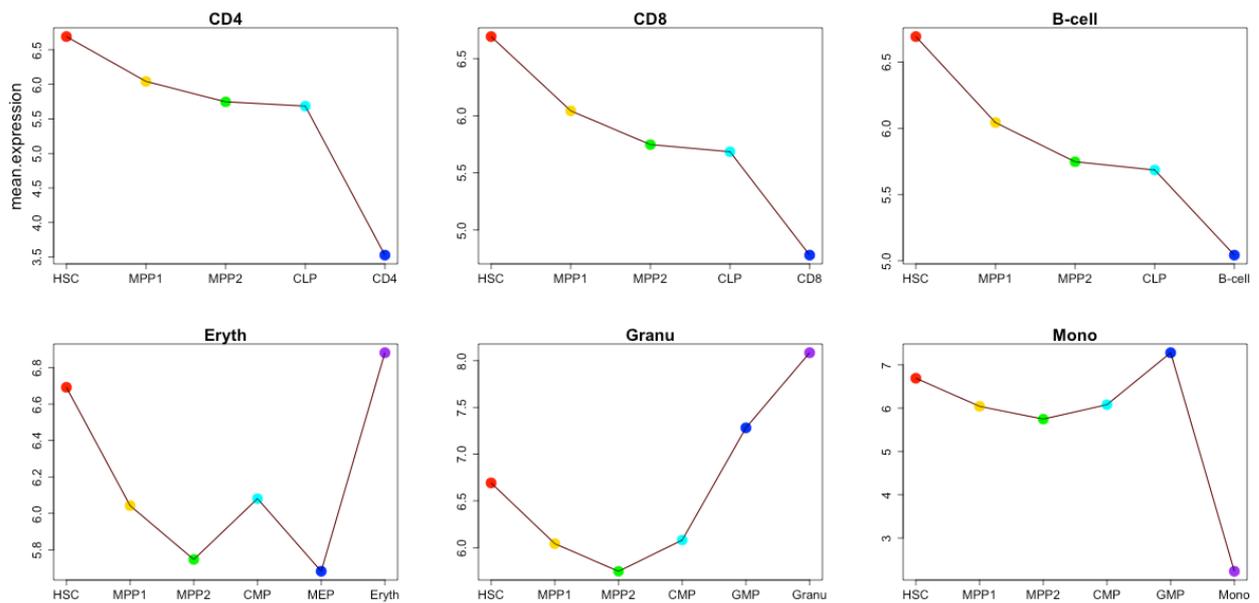

*Figure 10: Mean expression of stage-specific genes for the cells in each lineage for the six lineages CD4, CD8, B-cell, Eryth, Granu and Mono.*

Table S22 shows the number of lineage-specific correlated genes including the involved TFs. Additionally, it depicts the number of TFs that regulate the correlated genes inferred from the TRRUST database and the number of identified correlated genes that are targets of these TFs.

Tables S23-S28 contain the GO terms and KEGG pathways for the set of TFs that regulate the correlated genes mentioned in the second layer. GO terms such as GO:0045165, GO: 0001709, GO:0001708 annotated to cell fate commitment, cell fate determination and cell fate specification have been identified in the downstream analysis of almost all the lineage-specific TFs.

Table S29 shows the network statistics for the six lineages. As mentioned before, these networks consist of the derived TFs in the third layer and the target genes of the second layer. The network size lies between 81 and 272 nodes having 66 up to 293 interactions. Figure 11 illustrates the CD4 network constructed by the TFs and their target genes that overlap with the correlated genes in the second layer. The PathDevFate program highlighted genes (influencers colored red and connectors colored blue) that reside along the path to connect the influencers. Tables S29-S35 list these nodes for the six lineages including their roles and in-degree and out-degree. Table S36 displays the enriched GO terms and KEGG pathways for the set of nodes involved in the regulatory pathway of CD8 lineage. Among many terms related to cell differentiation and cell fate, GO: 0030217, that is annotated to the three involved genes Gata3, Ctnnb1 and Runx2, is specific to T cell differentiation.

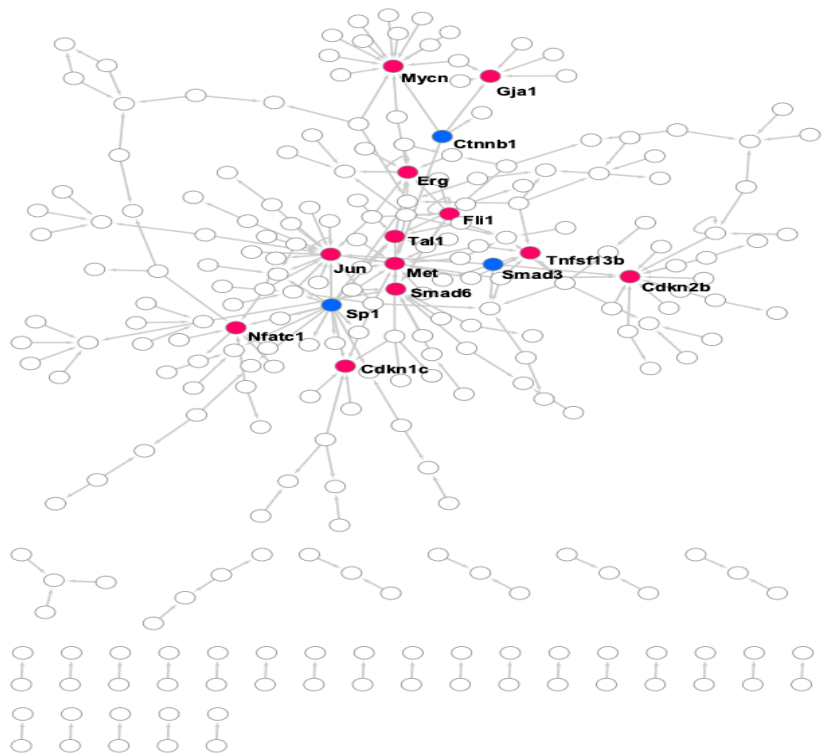

*Figure 11: The set of genes and TFs involved in the regulatory pathway. Influencers (red nodes) are the target genes that are regulated by more than 5 TFs. Connectors are TFs (blue nodes).*

## Discussion

In this work, we devised a pipeline for inferring a set of genes and TFs that drive the blood differentiation process controlling the cell fate decisions across a lineage starting at the stem cell stage and leading to a terminally differentiated stage. We started by identifying a set of genes and TFs having a particular stage-specific developmental expression pattern.

At the top level of this pipeline, we introduce a regulatory pathway in a gene-regulatory network of TFs and target genes taking into account the identified correlated genes and the TFs that regulate them. The regulatory pathway consists of a set of influencers that are regulated by multiple TFs and a set of connector TFs that connect them. The quality of this pathway depends on several points: First of all, the correlation threshold is a variable unless only perfectly correlated genes are to be considered. After that, the number of TFs that regulate these genes relies on the database(s) and the type of interaction which can be either experimentally confirmed (though likely not in the particular tissue investigated here) or predicted or both.

After all, the in-degree threshold for influencers is also a variable. The tighter threshold leads to a lower number of influencers but is not correlated with the size of the regulatory pathway. As shown in supplementary figure S1, in the lineages of CD8 and Granulocyte the number of connectors dramatically increases after a certain threshold. This observation indicates that those high-indegree influencers are very distant from each other and the algorithm needs to inject many connectors to connect them.

Principally, this work divides the identified genes and TFs into two groups. The first group describes the set of TFs that show the stage-specific developmental patterns and have a tendency to reach to the terminally differentiated state. The second group contains the set of TFs that regulate the set of genes and TFs which correlate with lineage-specific expression pattern. The regulatory pathway demonstrates a path that encompasses those correlated genes that are targeted by several TFs. This signifies the necessity of the gene to be involved in the process. Moreover, this pathway introduces a set of TFs to synchronize the activities of these influences in the lineage.

At this point, it is not very straight-forward to highlight the most important TFs as the number of TFs that are induced for connectivity highly depends on the number of influencers and the distance that these influencers have from each other in the network.

## Conclusion

In this work, we identified a set of genes and, from within this set, TFs that can be considered as potential biomarkers for the cell fate process during blood formation. To infer these candidates, we took as starting point the expression pattern of previously described global regulators in a blood lineage. Using this data, we identified stage-specific genes that are likely associated with the cellular differentiation based on correlated activity profiles. By combining the cell-specific expression pattern we reached to an integrated pattern specific to each lineage. Inferring the set of correlated genes and TFs that follow the lineage-specific expression pattern and incorporating the TFs that regulate the genes which have high correlation with the integrated pattern led to the identification of a regulatory subnetwork of TFs and their target genes. Nodes in these networks were finally prioritized using a newly developed "regulatory pathway" algorithm to find high-indegree genes and TFs by adding additional connector TFs.

All the nodes that reside along this path are suggested to be of a high priority for network function.

In this work, the set of TFs is prioritized in 4 layers. In the first layer, there are TFs that are mainly involved in the cellular differentiation process. The second layer consists of TFs that follow the integrated pattern of stage-specific expression pattern. TFs that regulate the correlated genes and TFs in the second layer constitute the candidate TFs in the third layer. Finally, the TFs that cooperatively regulate targets genes and connect high-indegree nodes (influencers) in the network of TFs in the third layer and the correlated genes and TFs in the second layer make the candidates in the fourth layer.

Enrichment analysis demonstrates that these biomarkers not only are involved in cell fate process but also in other developmental processes such as multicellular organism development etc. KEGG pathway analysis shows that these biomarkers can be potential targets for disease-related biomarkers such as leukaemia

In addition to the method which can be used as a computational approach to identify a regulatory pathway driving blood differentiation and also a set of genes and TFs that are introduced in four layers as potential biomarkers, the PathDevFate code can be used as a software to find the shortest path between a set of influencer nodes in the largest connected component where a user can set a threshold for the number of incoming edges.

## Data and code availability:

The data and code are available at https://github.com/ikmb/KeyDevelopmentalFate